\documentclass[12pt]{iopart}

\usepackage{iopams}  
\usepackage{graphicx}
\usepackage{color}
\usepackage{bm}
\usepackage{cite}

\begin{document}

\title{Controlled transport in chiral quantum walks on graphs}

\author{Yi-Cong Yu and Xiaoming Cai}


\address{Chinese Academy of Sciences Wuhan Institute of Physics and Mathematics,
Wuhan, Hubei, China}

\eads{\mailto{ycyu@wipm.ac.cn}}

\vspace{10pt}
\begin{indented}
\item[]\today
\end{indented}

\begin{abstract}

We investigate novel transport properties of chiral continuous-time quantum walks (CTQWs) on graphs. 
By employing a gauge transformation, we demonstrate that CTQWs on chiral chains are equivalent to those on non-chiral chains, but with additional momenta from initial wave packets.
This explains the novel transport phenomenon numerically studied in [New J. Phys. \textbf{23}, 083005(2021)]. 
Building on this, we delve deeper into the analysis of chiral CTQWs on the Y-junction graph, introducing phases to account for the chirality. The phase plays a key role in controlling both asymmetric transport and directed complete transport among the chains in the Y-junction graph.
We systematically analyze these features through a comprehensive examination of the chiral continuous-time quantum walk (CTQW) on a Y-junction graph. Our analysis shows that the CTQW on Y-junction graph can be modeled as a combination of three wave functions, each of which evolves independently on three effective open chains. By constructing a lattice scattering theory, we calculate the phase shift of a wave packet after it interacts with the potential-shifted boundary. Our results demonstrate that the interplay of these phase shifts leads to the observed enhancement and suppression of quantum transport.
The explicit condition for directed complete transport or $100\%$ efficiency is analytically derived. 
Our theory has applications in building quantum versions of binary tree search algorithms. 

\end{abstract}

\vspace{2pc}

\submitto{\NJP}

\maketitle

%
%
%
%
%

\section{Introduction} \label{secIntroduction}

Quantum walks, the quantum mechanical counterparts of classical random walks, are advanced tools for building quantum algorithms that have been recently shown to constitute a universal model of quantum computation \cite{aharonov1993, venegas-andraca2012,manouchehri2014}. 
For example, quantum search algorithms \cite{grover1997,childs2004} and quantum decision tree algorithms \cite{farhi1998,magniez2011} have demonstrated the quantum superiority by significantly reducing the time complexity in comparison to classical algorithms.
The essential physical bases of them are the discrete-time quantum walk (DTQW) and the continuous-time quantum walk (CTQW) respectively. 
The key distinction between DTQW and CTQW lies in the timing applied in their evolution operators, with the former being discrete and the latter continuous \cite{venegas-andraca2012, mulken2011}. Despite this difference, they have been demonstrated to be equivalent in the long-time limit \cite{childs2010}.
Besides being a theoretical tool for studying quantum computation, the CTQW also has become a standard model for understanding quantum transport phenomena on graphs \cite{childs2003,kendon2007,dimolfetta2021,todtli2016}, since in the  context of quantum mechanics the CTQW is equivalent to the dynamical problem of a given Hamiltonian\cite{mulken2011}. 
It is believed that understanding transports in quantum walks is the key to developing more robust communication networks, more effective energy transmissions, and more reliable information processing devices \cite{chen2017,manzano2018,sett2019,xiang2013,mohseni2008}.
A disadvantage of the classic CTQW is the inability to model or analyze directed graphs, due to the unitary condition imposed by fundamental quantum mechanical postulates \cite{izaac2017}.
Although biased transport can be realized by time-dependent controlling or by introducing non-Hermiticity \cite{chen2017,manzano2018,rosa2020}, achieving the directional bias in quantum walks while keeping the unitary of time evolution  proved challenging. 
It was reported until recently \cite{khalique2021} that a delicate Hamiltonian with varying phases on edge-weights could achieve directed transports on path and cycle graphs.
This type of quantum walks is called the chiral quantum walk \cite{lu2016} with the introduced phases breaking the time-reversal symmetry. 
In the past decade, the chiral quantum walk has attracted lots of attentions \cite{lu2016,lodahl2017,vahedi2022,jen2020,javaherian2017}.
Research has shown that chirality can lead to novel physical phenomena \cite{sett2019,saglam2021}, inspiring further exploration both theoretically \cite{marsh2021,zhou2021,burgarth2013,cameron2014,godsil2010,izaac2017,maciel2020,mares2020,gou2020} and experiments \cite{wang2020,wu2020,roushan2017}.
In the presence of chirality, asymmetric and non-reciprocal transports were experimentally demonstrated in quantum circuits \cite{lu2016}, quantum optics \cite{lodahl2017}, and optical waveguides \cite{javaherian2017}.
Asymmetric spin transport \cite{vahedi2022} and steady state transport \cite{jen2020} were proposed to emerge in chiral spin and atomic chains.
On graphs, completely suppressed flow and fast entanglement transfer have been proposed  \cite{sett2019,saglam2021}.
In particular, it was demonstrated in \cite{zimboras2013} that the transport efficiency in quantum walks on graphs could be greatly enhanced by controlling the chirality, and even the direction of transport can be controlled.
However, previous research has only achieved partial controllability or partial transport on a selective direction. 

In this paper, we examine chiral continuous-time quantum walks (CTQWs) on graphs with branching structures, with a focus on the Y-junction graph and its controllable transport properties.
Unlike previous limited capabilities, the direction of transport and the proportions of transported wave packets on different branches can be fully controlled by adjusting the phase of the edge-weights.
By selecting specific phases, it is possible to achieve complete transport in a desired direction or  $100\%$ efficiency.
To shed light on these novel transport phenomena, we first analytically demonstrate that the chiral CTQW on the Y-junction graph is equivalent to a superposition of three wave packets that evolve on three effective open chains with potential-shifted boundaries, respectively.
Next, by constructing the scattering theory on a lattice, we determine the phase-shift of a wave packet scattered by a potential-shifted boundary and reveal that the enhancement and suppression of the superposition of three phase-shifts precisely underlies the (a)symmetric transport and the directed complete transport.
We also analytically derive the explicit conditions for directed complete transport.
In addition, we demonstrate that our theory has applications in the development of quantum binary tree search algorithms and can be experimentally tested in various setups.

The paper is organized as follows. In Sec.\ref{secDirected}, we study the chiral CTQW on an open chain and explain the novel directed transport phenonenon analytically. In Sec.\ref{secTriangleSwitch} we study the chiral CTQW on the Y-junction graph, firstly show examples of the controlled transport phenonena by numerical results, and then solve the Hamiltonian analytically. Based on the results in Sec.\ref{secTriangleSwitch} we build the scattering theory on the Y-junction graph in Sec.\ref{secScattering} to find the condition for the controlled transport. We also show some more examples in the appendix to proof the universality and  robustness of the controlled transport.

\section{Chiral quantum walk and directed transport} \label{secDirected}

The Hamiltonian for the chiral CTQW (or simply the chiral quantum walk) of a particle (walker) on a graph with $N$ sites (or vertices) is \cite{zimboras2013}
\begin{equation}
	\hat{H} = \sum_{\langle m, n\rangle} J_{nm} \mathrm{e}^{\mathrm{i}\theta_{nm} }\vert n \rangle \langle m \vert
	+ J_{nm} \mathrm{e}^{-\mathrm{i}\theta_{nm} }\vert m \rangle \langle n \vert .
	\label{eqHamiltonian}
\end{equation}
$\vert k\rangle$ is the orthonormal Wannier base localized at site $k = 1,2, \cdots,  N$. 
$\langle \cdot , \cdot \rangle$ denotes the summation over all adjacent sites. 
$J_{nm}$ are real numbers describing edge-weights or hopping amplitudes between neighbors. 
$\theta_{nm} \in [0, 2\pi)$ are phases of the hoppings  breaking the time-reversal symmetry \cite{lu2016, zimboras2013}. 
The time evolution of the system is governed by the unitary operator $\hat{U}(t) = \exp(-\mathrm{i}\hat{H}t)$.
For simplicity, we consider a uniform hopping amplitude, setting $J_{nm} = J = 1$.
In the absence of time-reversal symmetry ($\theta_{nm} \neq 0, \pi$), a walker can move in a preferred direction, while its movement in other directions is suppressed, which makes the transport time-asymmetric or chiral. 
%
%
We first consider the chiral quantum walk on a chain which was studied in \cite{khalique2021}. 
By choosing phases that scale linearly with the coordinate of sites, it was showed numerically that a directed transport, instead of expansion or asymmetric transport, of a Gaussian wave package with initial zero momentum can be achieved. 
Here, we give an analytical explanation of the counter-intuitive phenomenon. 
The adjacency matrix of chain (matrix representation of the Hamiltonian in Wannier basis) can be written as
\cite{khalique2021}
\begin{equation}
\bm{A} = \left(
\begin{array}{cccccc}
 0 & 1 &  &  &  &\\
1  & 0 & \mathrm{e}^{\mathrm{i}\frac{\pi}{N-2}} & &  &\\
 & \mathrm{e}^{-\mathrm{i}\frac{\pi}{N-2}} & 0 & \mathrm{e}^{\mathrm{i}\frac{2\pi}{N-2}} &  & \\
 &  &  \mathrm{e}^{-\mathrm{i}\frac{2\pi}{N-2}} & 0 & \ddots & \\
 & & & \ddots & \ddots & \mathrm{e}^{\mathrm{i}\pi} \\
 &  & &  & \mathrm{e}^{-\mathrm{i} \pi} & 0 
\end{array}
\right).
\label{eqA}
\end{equation}
It can be unitarily transformed into a simple tridiagonal matrix $\bm{D} = \bm{C A C}^\dagger$ with elements $D_{mn} = \delta_{m, n - 1} + \delta_{m,  n + 1}$.
The diagonal and unitary matrix $\bm{C}$ corresponds to a gauge transformation, which keeps the local density invariant \cite{lu2016}. 
Its elements are $C_{mn} = \delta_{m,n}\exp(-\mathrm{i} \alpha_n) $ with 
\begin{equation}
\alpha_n = -\frac{\pi}{2(N-2)} (n-2)(n-1).
\label{eqAlpha}
\end{equation}
Generally, the Hamiltonian, which governs chiral quantum walks on chains, can be written as
$\hat{H}_c 
= \sum_{n} J_{n} \mathrm{e}^{\mathrm{i}\theta_{n} }\vert n \rangle \langle n+1  \vert
+ \mathrm{h.c.}$. 
By introducing a gauge transformation $\vert n \rangle  = \mathrm{e}^{\mathrm{i} \alpha_n} \vert n \rangle ^\prime$ with phases satisfying $\alpha_n - \alpha_{n+1} = \theta_n$, the Hamiltonian $\hat{H}_c$ is turned into a phase-free form 
$\hat{H}'_c = \sum_{n} J_{n} \vert n \rangle^\prime \langle n+1 \vert ^\prime+ \mathrm{h.c.} $.
Meanwhile, a given time-reversal symmetric initial state $\vert \psi_0 \rangle = \sum_n \phi^0_n \vert n \rangle$ is transformed into $\vert \psi_0 \rangle = \sum_n \phi^0_n \mathrm{e}^{\mathrm{i}\alpha_n} \vert n \rangle^\prime$. 
%
%
%
In a word, through a gauge transformation, the chiral quantum walk, which is governed by a time-reversal symmetry broken Hamiltonian and from a symmetric initial state, is equivalent to the one for a symmetric Hamiltonian but a symmetry broken initial state. 
For the initial state discussed in \cite{lu2016, khalique2021}, which was deliberately chosen as a Gaussian wave package locating at the center of the chain, the phase difference between adjacent sites near the center after the gauge transformation according to (\ref{eqAlpha}) with taking the limit $N \to \infty, n \to N/2$ is given by 
\begin{equation}
\lim_{n = \frac{N}{2} \to \infty}\alpha_{n+1} - \alpha_n = -\frac{\pi}{N-2}(n-1) = -\frac{\pi}{2} + O(\frac{1}{N}),
\end{equation}
which means that the intial wave package obtains a momentum $-\pi/2$ after the gauge transformation. 
If we  consider the low energy physics, i.e. the wave function varies slowly with $n$, we can replace the basis $\vert n \rangle$ by $\vert n \rangle \to \Psi^\dagger(x=n)\vert \Omega \rangle$, here $\Psi^\dagger(x)$ is the field creating operator, $\vert \Omega \rangle$ denotes the vacuum, then the Hamiltonian described by the operator $\bm{D}$ can be represented by $\bm{D} \vert n \rangle = \vert n-1 \rangle + \vert n+1 \rangle = (2 \Psi^\dagger(x) + \frac{\partial^2\Psi^\dagger(x)}{\partial x^2} ) \vert \Omega \rangle = \bm{D} \Psi^\dagger(x) \vert \Omega \rangle$. Dropping the non-important constant in $\bm{D}$ which only leads to a global phase, the Schr\"{o}dinger equation of amplitudes $\phi_n(t)$ is mapped to the Cauchy equation for wave propagation $\mathrm{i} \frac{\partial \phi(x, t)}{\partial t} = \frac{\partial^2 \phi(x, t)}{\partial x^2}$, where $x$ and $\phi(x, t)$ are the coordinate and wave function respectively.
The corresponding initial state after gauge transformation is $\phi(x,0) = g(x)\exp(\mathrm{i} k_0 x)$, where $g(x)\propto e^{-x^2/\sigma^2}$ refers to the initial wide Gaussian function, and $k_0 = -\pi/2$ is the initial momentum of wave package which originates from the phase difference at the center of chain. 
The Cauchy equation can be solved by the Fourier transformation and results in
\begin{equation}
\phi(x, t) = \int \mathrm{d} p \, \tilde{\phi}(p)  \mathrm{e}^{-\mathrm{i}(\epsilon_p t - px)}   \approx \mathrm{e}^{-\mathrm{i}(\epsilon_{k_0}-k_0 x)} \int \mathrm{d} p \, \tilde{\phi}(p) \mathrm{e}^{-\mathrm{i}(p-k_0)(\epsilon^\prime_{k_0} t -x )},
\label{eqCauchy}
\end{equation}
where $\tilde{\phi}(p)$ is the Fourier transformation of the initial wave function $\phi(x,0)$, and $\epsilon_p \sim 2\cos p$ is the spectrum of matrix $\bm{D}$ in the continuum limit.  This result can be obtained by first approximate the operator $\bm{D}$ by the periodic one $\bm{D}^\mathrm{per}$ (because we are dealing with the transport problem in the middle of the chain, the difference caused by boundary condition can be neglected reasonably) with $D^\mathrm{per}_{mn} = \delta_{m, n - 1} + \delta_{m,  n + 1} + \delta_{m, 1}\delta_{n, N} + \delta_{m,N}\delta_{n,1}$, and then diagonalize this matrix by unitary matrix $U_{mn} = N^{-1/2} \exp{(\mathrm{i}\frac{mn}{N})}$, obtaining finally the eigen values $2\cos \frac{2\pi k}{N}$ for $k = 0, 1, \cdots, (N-1)$.
The approximation made above is based on the fact that $\tilde{\phi}(p)$ is narrow Gaussian and centers on $k_0$. 
From (\ref{eqCauchy}) we learn that group velocity of the wave package is $v_\mathrm{g} = \epsilon^\prime_{k_0}$. 
Furthermore, obtained by partial integration, deformation of the propagating wave package is determined by the second derivative of spectrum $\epsilon_p$, which is $\epsilon^{\prime\prime}_{p} \sim -2\cos(p)$.
The deformation is absent at $p=k_0 = -\pi/2$, where the second derivative $\epsilon^{\prime\prime}=0$. 
This is precisely the physical reason for the high fidelity of wave package transport numerically studied in \cite{khalique2021}.

\begin{figure}[ht]
	\begin{center}
	\includegraphics[width=0.8\textwidth]{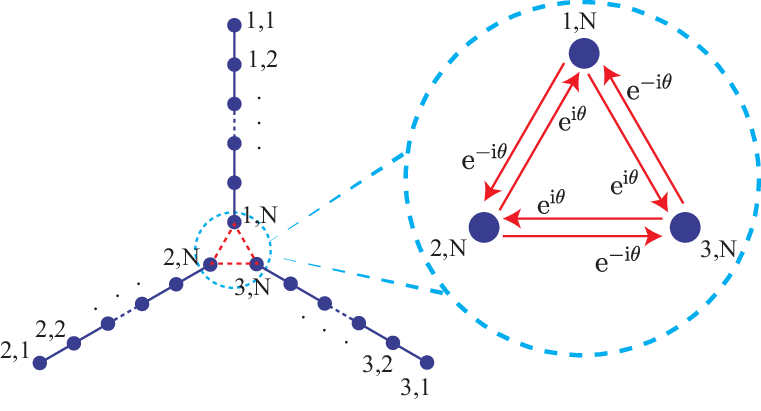}
	\caption{Schematic of the Y-junction graph.
		Three non-chiral chains with unit edge-weights are connected by their end sites with a triangle.
		In the presence of phase $\theta$, edge-weights on triangle make the graph chiral.}
	\label{figSchematic1}
	\end{center}
\end{figure}

\section{Controlled transport on Y-junction graph} \label{secTriangleSwitch}

The Y-junction graph, in which the phases on the edge-weights can not be gauged away as in Section \ref{secDirected} due to the loop nature, can bring novel phenomena in quantum walks such as directed currents \cite{roushan2017}, non-reciprocal transmission \cite{gou2020}, and quantum transport enhancement \cite{zimboras2013}. 
It was paid close attention recently both in theories \cite{khalique2021,zimboras2013,saglam2021,sett2019} and experiments \cite{wang2020,papadopoulos2000,tang2018,golshani2014,tkachenko2012}. 
The Y-junction graph can be realized in experiments by high-dimensional photonic quantum states, multiple concatenated interferometers, or dimension-dependent losses \cite{wu2020}. 

In the following, we study chiral quantum walks on the Y-junction graph shown in Figure \ref{figSchematic1}, which is a combination of three chains and a triangle. 
The three phase-free (non-chiral) chains are modeled by the Hamiltonian 
\begin{equation} 
\hat{H}^l_\mathrm{c} = J\sum_{n=1}^{N-1} \Big(\vert l, n \rangle\langle l, n+1  \vert
+ \mathrm{h.c.}\Big),
\label{eqHamc}
\end{equation}
where $l=1,2,3$ denotes the index of chain, and $\vert l, n \rangle$ refers to the Wannier base locating at the $n$-th site of the $l$-th chain.  
These three chains are connected by a triangle which is described by the Hamiltonian 
\begin{equation}
	\hat{H}_\mathrm{t}=J\mathrm{e}^{\mathrm{i}\theta}\Big(\vert 1, N \rangle\langle 2, N  \vert+\vert 2, N \rangle\langle 3, N  \vert+\vert 3, N \rangle\langle 1, N \vert\Big)+\mathrm{h.c.}.
\label{eqHamt}
\end{equation}
The tunable phase $\theta$ introduced in (\ref{eqHamt}) breaks the time-reversal symmetry and makes quantum walks on the graph chiral.  
The full Hamiltonian governing chiral quantum walks on the Y-junction graph is 
\begin{eqnarray}
\hat{H}=\hat{H}_\mathrm{t} + \sum_{l=1,2,3}\hat{H}^l_\mathrm{c} . 
\label{eqFullHam}
\end{eqnarray}

The initial states is crucial in the study of quantum walks, as it significantly impacts the transport properties of the walkers \cite{venegas-andraca2012,kempe2003,mulken2011}.
Two commonly used initial states have received extensive attention and been thoroughly studied: a Gaussian wave packet centered on a specific site in the graph, commonly used in quantum decision tree algorithms \cite{childs2009,farhi1998,mulken2011,manouchehri2008}, and a square wave packet uniformly distributed across multiple consecutive sites, which is favored in quantum search algorithm research \cite{grover1997,childs2004,magniez2011,su2019,sanchez-burillo2012}.
Given that quantum walks from these two typical initial wave packets exhibit similar major behaviors in our study, we will focus on the transport properties of the Gaussian wave packet in detail,
and present the square wave packet case in the Appendix.
%

We select the initial Gaussian wave packet to be located on the first chain, with a given initial momentum. It is described by $|\psi_0\rangle=(2\pi\sigma^2)^{-1/2}\sum_n\mathrm{e}^{-(n-n_0)^2/2\sigma^2-\mathrm{i}k_0n}|1,n\rangle$, where $n_0$ and $\sigma$ represent the center and width of the initial wave packet, respectively. $k_0$ is the initial momentum, which can also be introduced through the gauge transformation described in Section \ref{secDirected}.
For simplicity, we have set $n_0=N/2$, $\sigma=N/\sqrt{32}$, and focused on the initial momentum of $k_0=\pi/2$, unless specified otherwise.

\begin{figure}[tb]
	\begin{center}
	\includegraphics[width=0.9\textwidth]{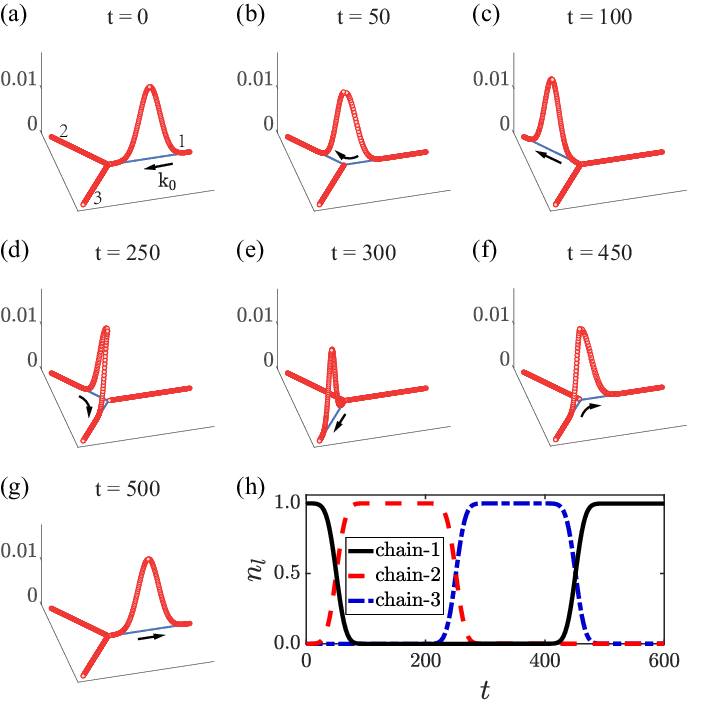}
	\caption{Snapshots of the chiral quantum walk on the Y-junction graph (figure \ref{figSchematic1}) with $N = 200$ and $\theta = \pi/6$. 
		The $Z$-axis in (a-g) refers to the local density.	
		(a) The initial Gaussian wave package, centering on the center of chain-1 and with the initial momentum $k_0 = \pi/2$.  	
		(b) The wave package is transported from chain-1 to chain-2 while leaving chain-3 completely untouched. 	
		(c) The transported wave package on chain-2.
		(d) Reflected by the boundary of chain-2, the wave package is transported to chain-3. 	
		(e) The transported wave package on chain-3.	
		(f) The wave package is transported from chain-3 to chain-1. 
		(g) The wave package is back on chain-1. 	
		(h) Time evolutions of total densities of particles on different chains, defined by  (\ref{eqDefChainDens}).			
		Note that during the propagation, group velocity of the wave package is a constant [$v_\mathrm{g} = 2\sin (\pi/2) = 2$] and its envelope is also invariant.
	 }
	\label{figDynamic1}
	\end{center}
\end{figure}

\subsection{Phase-controlled transport} \label{secResult}

\begin{figure}[t]
	\begin{center}
	\includegraphics[width=0.9\textwidth]{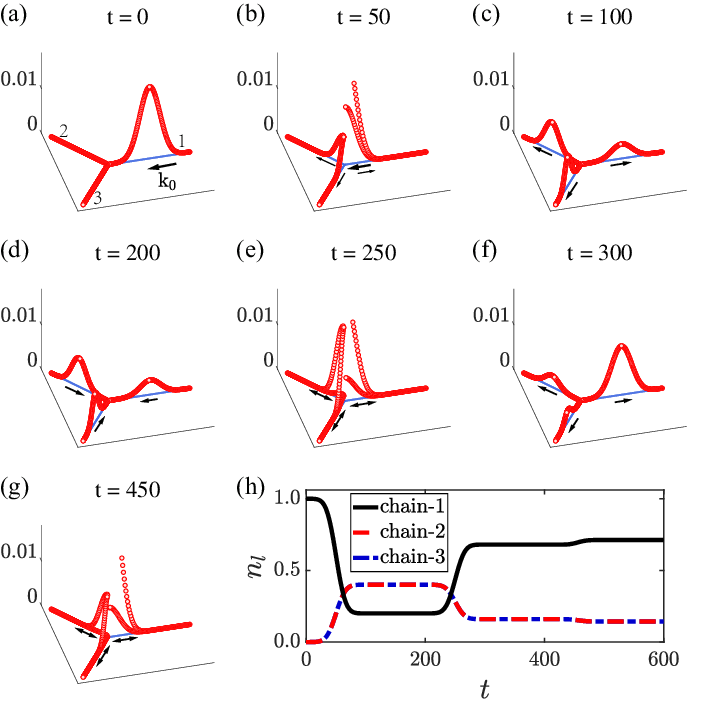}
	\caption{(a-g) Snapshots of the chiral quantum walk, from the same initial wave package as in Figure \ref{figDynamic1}, and on the Y-junction graph with $N = 200$ but a different $\theta = 0$. 
		After the first collision with the Y-junction, the wave package splits into three parts. 
		Equal two of them are transported onto the right- and left-hand side chains, respectively, while the other part is reflected back. 
		(h) Time evolutions of total densities of particles on different chains. } 
	\label{figDynamic2}
	\end{center}
\end{figure}

\begin{figure}[tb]
	\begin{center}
		\includegraphics[width=1.00\textwidth]{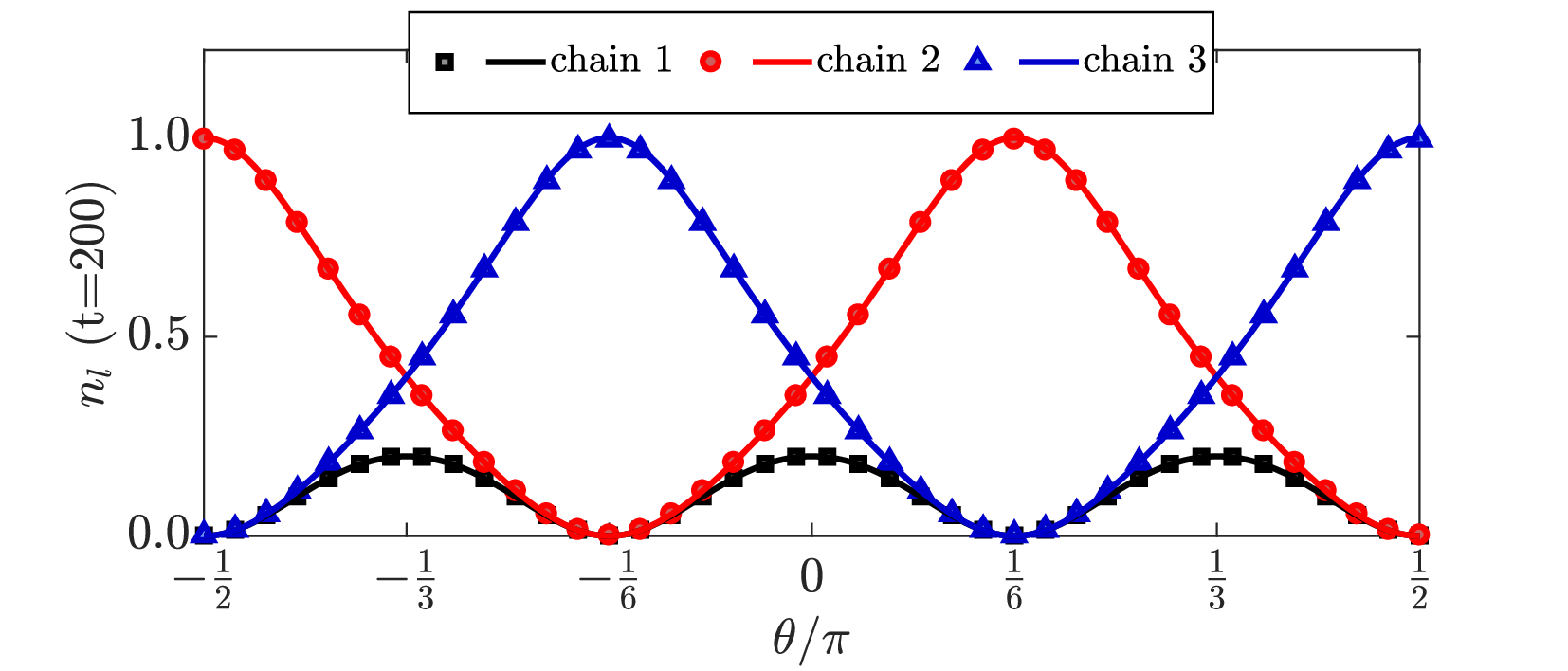}
		\caption{Total densities of particles on different chains $n_l$ vs. phase $\theta$,  after the first collision of the wave package with the Y-junction. 
			Parameters are the same as in Figures \ref{figDynamic1} and \ref{figDynamic2}.  
			Solid lines refer to analytical results, while marked points correspond to numerical simulations. 
			Directed complete transport occurs at critical points $\theta = \cdots, -\pi/2, -\pi/6, \pi/6, \pi/2, \cdots$ [see also (\ref{eqRelationKTheta})]. }
		\label{figTheta}
	\end{center}
\end{figure}

The time evolution of the initial wave packet can be computed numerically by computing $\exp(-\mathrm{i}t\hat{H})|\psi_0\rangle$ after performing exact diagonalization of the Hamiltonian in (\ref{eqFullHam}). In Figures \ref{figDynamic1}(a-g), we display snapshots of the time evolution of the density distribution for the chiral quantum walk on the Y-junction graph with a phase of $\theta = \pi/6$. The corresponding movie is included in the supplemental material.

Starting from the first chain, the Gaussian wave packet travels towards the Y-junction. Upon being scattered by the junction, it is fully transported to the second chain as if the junction was not present. Then, after being reflected by the end of the second chain and scattered by the junction again, it is fully transported to the third chain. Over time, the wave packet is fully transported to chains 2, 3, 1, 2, 3, and so on in succession. In short, after being scattered by the junction, the wave packet is always fully transported to the chain on its right-hand side.

The period of this cyclic propagation is determined by the number of sites divided by the group velocity of the wave packet, $v_\mathrm{g} = 2\sin(\pi/2) = 2$. Additionally, the width of the wave packet remains nearly constant during the propagation due to the vanishing second derivative of the spectrum at $k_0=\pi/2$.
To further quantify the transport of the wave package, we compute the density of particles on the $l$-th chain, which is defined by 
\begin{equation}
n_l (t)= \sum_n \vert \phi_{l,n}(t) \vert^2.
\label{eqDefChainDens}
\end{equation}
$\phi_{l,n}(t)$ is the amplitude of the evolving wave function at site $n$ of the $l$-th chain. 
In Figure \ref{figDynamic1}(h), the evolution of the total particle probability densities on each chain for a chiral quantum walk on a Y-junction graph with a phase of $\theta=\pi/6$ is shown. The periodic propagation and complete transport of the wave package among chains are easily observed.

In contrast, Figure \ref{figDynamic2}(a-g) displays snapshots of the density distribution evolution for a chiral quantum walk on a Y-junction graph with a phase of $\theta=0$. The corresponding evolution of the total particle probability densities on each chain is presented in Figure \ref{figDynamic2}(h). Since the governing Hamiltonian is time-reversal symmetric when $\theta=0$, the quantum walk is significantly different from the one shown in Figure \ref{figDynamic1}. After being scattered by the junction, part of the wave package splits equally onto the second and third chains, while the other part is reflected back. Interference patterns form as incoming and reflected wave packages collide with boundaries or the junction. The superposition of the split and reflected wave packages from different chains results in a complicated behavior in the total particle probability densities, losing its periodicity at least for a short time.

The non-zero $\theta$ breaks the time-reversal symmetry of the system, resulting in asymmetric transport. The phase $\theta$ determines the direction of complete transport, with $\theta=\pi/6$ corresponding to complete transport onto chain-2, and $\theta=-\pi/6$ to complete transport onto chain-3. In the case of $\theta=0$, the wave package splits equally onto chains-2 and 3, with the strongest reflection. The total densities of particles on chains are periodic functions of the phase $\theta$, with a common period $2\pi/3$. These results show that the phase $\theta$ can be used to control the reflection, transport, and complete transport of the wave package in a chiral quantum walk on a Y-junction graph.




\subsection{Spectrum and eigenstates} \label{secEigen}

The time evolution of a wave package from the initial state $|\psi_0\rangle$ is described by $\vert\psi(t)\rangle=\mathrm{e}^{-\mathrm{i}\hat{H}t}\vert\psi_0\rangle$. 
With eigen-decomposition, it can be rewritten as
\begin{equation}
	\vert \psi(t) \rangle  = \sum_{\alpha} \mathrm{e}^{-\mathrm{i}\epsilon_\alpha t}
	\vert \alpha \rangle \langle \alpha \vert \psi_0 \rangle ,
	\label{eqPsit}
\end{equation}
where $\vert \alpha \rangle$ is the $\alpha$-th eigenstate of $\hat{H}$ and $\epsilon_\alpha$ is its eigenenergy. 
Given that the time evolution governed by a time-independent Hamiltonian is determined by its stationary properties \cite{abrikosov2012,taylor1972,newton2002}, we first solve the eigenvalue problem in this subsection.

Notice that the Hamiltoian (\ref{eqFullHam}) is with the $Z_3$ symmetry, thus the Hilbert space can be split into three independent ones, and we will show that each of the subspace can be described by an effective Hamiltonian with parameter $\theta$ and  solved analytically. To achieve this idea, we lable 
the eigenstate of $\hat{H}$ is described by amplitudes $\phi_{l,n}$, with $n$ and $l$ indexes of site and chain respectively. 
We denote eigenenergy of the state by $\epsilon\equiv\lambda+\lambda^{-1}$ and employ the open boundary condition for each chain. 
The eigenequation results in a recurrence relation between amplitudes $\phi_{l,n}/{s_n}=\phi_{l,m}/{s_m}$, where $s_m\equiv\lambda^m-\lambda^{-m}$. 
The remaining unknown amplitudes $\phi_{1,N}$, $\phi_{2,N}$, and $\phi_{3,N}$ are determined by eigenequation at the junction
\begin{equation}
	\left\{
	\begin{array}{c}
		\epsilon \phi_{1,N} = \frac{s_{N-1}}{s_N} \phi_{1,N} 	+ \mathrm{e}^{\mathrm{i}\theta} \phi_{2,N} + \mathrm{e}^{-\mathrm{i}\theta} \phi_{3,N}, \\
		\epsilon \phi_{2,N} = \frac{s_{N-1}}{s_N} \phi_{2,N} + \mathrm{e}^{\mathrm{i}\theta} \phi_{3,N} + \mathrm{e}^{-\mathrm{i}\theta} \phi_{1,N}, \\
		\epsilon \phi_{3,N} = \frac{s_{N-1}}{s_N} \phi_{3,N} 	+ \mathrm{e}^{\mathrm{i}\theta} \phi_{1,N} + \mathrm{e}^{-\mathrm{i}\theta} \phi_{2,N},
	\end{array}
	\right.
	\label{eqM3eigen}
\end{equation}
with $N$ the number of sites in each chain. 
Introducing the adjacency matrix for the junction
\begin{equation}
	\bm{M} =	\left(
	\begin{array}{ccc}
		0 & \mathrm{e}^{\mathrm{i}\theta} & \mathrm{e}^{-\mathrm{i}\theta} \\
		\mathrm{e}^{-\mathrm{i}\theta} & 0 & \mathrm{e}^{\mathrm{i}\theta} \\
		\mathrm{e}^{\mathrm{i}\theta} & \mathrm{e}^{-\mathrm{i}\theta} & 0
	\end{array}
	\right),
	\label{eqM0}
\end{equation}
one can easily verify that the vector $\bm{\Phi}\equiv[\phi_{1,N}, \phi_{2,N}, \phi_{3,N}]^T$ is an eigenvector of $\bm{M}$. 
Diagonalizing the matrix $\bm{M}$, we obtain eigenvalues
\begin{equation}
	\omega_\nu=2\cos(\frac{2\pi}{3}\nu+\theta),\quad \nu=1,2,3,
	\label{eqOmega}
\end{equation}
and eigenvectors 
$\phi^\nu_{l,N}=\frac{1}{\sqrt{3}}\exp(\mathrm{i}\frac{2\pi}{3}\nu l)$.
$\lambda$ and then eigenenergies $\epsilon$ are determined by $\omega_\nu=\epsilon-\frac{s_{N-1}}{s_N}$, or
\begin{equation}
	\frac{\lambda^{N+1}-\lambda^{-1-N}}{\lambda^{N}-\lambda^{-N}}=\omega_\nu.
	\label{eqPolyEq}
\end{equation}
There are $3N$ roots $\lambda_{\nu\eta}$ satisfying both $|\lambda| \geq 1$
and $\mathrm{arg}\lambda \in [0, \pi)$, $N$ roots for each $\nu$. 
From the recurrence relation, we obtain eigenstates of $\hat{H}$
\begin{equation}
	\phi^{\nu\eta}_{l,n}=c_{\nu\eta}\frac{1}{\sqrt{3}}\exp(\mathrm{i}\frac{2\pi}{3}\nu l)(\lambda^n_{\nu\eta}-\lambda^{-n}_{\nu\eta}),\quad \nu = 1,2,3, \, \eta=1,2,...N,
\label{eqWaveFunc}
\end{equation}
with $c_{\nu\eta}$ unimportant normalization coefficients. 
Notice that for every eigenstate, amplitudes on different chains only differ by global phases $0$, $2\pi/3$, and $-2\pi/3$. 
Thus, an eigenstate on the Y-junction graph is totally determined by its wave function on one chain, say the first chain, and density distributions of it on different chains are the same. 
Based on this observation, we transform the eigenvalue problem of $\hat{H}$ into the ones of three effective Hamiltonians $\hat{H}_{\mathrm{eff}}^\nu$ on a single chain, where
\begin{equation}
	\hat{H}_{\mathrm{eff}}^\nu = J\sum_{n=1}^{N-1} \Big(\vert n \rangle\langle n+1  \vert
	+ \mathrm{h.c.}\Big)+\omega_\nu\vert N \rangle\langle  N  \vert, \quad\nu=1,2,3.
\label{eqHamEff}
\end{equation}
The eigenvalues $\omega_\nu$ of the adjacency matrix $M$ in (\ref{eqM0}) turn into potential shifts on the edge site in $\hat{H}_{\mathrm{eff}}^\nu$. The solution of the original Hamlitonian (\ref{eqFullHam}) can be obtained by solving these three effective Hamiltonian (\ref{eqHamEff}) independently. 

\begin{figure}[tb]
	\begin{center}
		\includegraphics[width=1.00\textwidth]{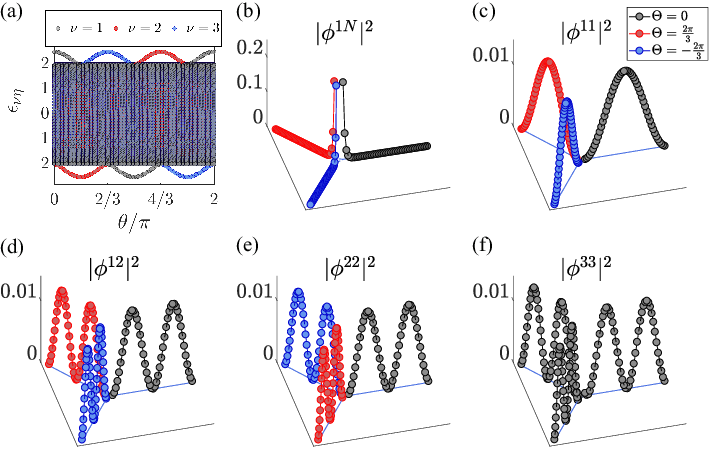}
		\caption{Spectra and typical eigenstates of the Hamiltonian in  (\ref{eqFullHam}).
			(a) The spectrum vs. the phase $\theta$. 
			Different colors refer to $\nu = 1, 2, 3$, for different subspaces or effective chains [equation (\ref{eqHamEff})]. 
			Edge states appear above and below the continuous energy band. 
			(b-f) Distributions of typical eigenstates, obtained from (\ref{eqWaveFunc}).
			Different colors correspond to different phases on three chains.
			The eigenstate $\phi^{1N}$ shown in (b) is localized at the Y-junction, while $\phi^{12}$, $\phi^{22}$, and $\phi^{33}$ are extended, differing only by relative phases among different chains.
			Here lattice size is $N = 50$, and the phase $\theta=\pi/6$.}
		\label{figSpecWave}
	\end{center}
\end{figure}

To obtain wave functions on a single chain, we first define $\lambda = \mathrm{e}^{\mathrm{i}\zeta}$. 
Then (\ref{eqPolyEq}) is simplified to  $\sin(N+1)\zeta / \sin N\zeta = \omega_\nu $. 
The left-hand side of it monotonically decreases as the phase $\zeta$ increases, in every interval $(\frac{\pi}{N}(m-1), \frac{\pi}{N}m)$, $m = 1, 2, \cdots, N$, and it diverges at $\zeta = \frac{\pi}{N}m$ for $m = 2, 3, \cdots, N-1$. 
This leads to the fact that when $\vert \omega_\nu \vert \leq (N+1)/N$,  $N$ real roots $\zeta$ can be found in the above equation, one in each interval, while when 
$\vert \omega_\nu \vert > (N+1)/N$, one imaginary root exists and the remaining $N-1$ roots are real. 
Given that the wave function is of the form $\vert \psi \rangle \sim \sin(n \zeta)$, the real $\zeta$ corresponds to the extended state while the imaginary $\zeta$ corresponds to the localized one.  
Spectra and distributions of typical eigenstates are presented in Figure \ref{figSpecWave}. 

\subsection{Decomposition of the time evolution and directed transport} \label{secDecomposition}

We first define three operators $P_l$ by $P_l|n\rangle = |l,n\rangle$, which map the Wannier basis of the chain to the ones of the Y-junction graph. 
They satisfy $P_l^\dagger P_{l'}=\delta_{ll'}$. 
Then according to  (\ref{eqWaveFunc}), eigenstates of $\hat{H}$ can be rewritten as
\begin{eqnarray}
\vert \psi^{\nu\eta}\rangle&=&\sum_{l,n}\phi^{\nu\eta}_{l,n}\vert l,n\rangle
=\frac{1}{\sqrt{3}}\sum_l\sigma^{\nu l}P_l \vert \chi^{\nu\eta}\rangle,\\
|\chi^{\nu\eta}\rangle&=&c_{\nu\eta}\sum_n(\lambda^n_{\nu\eta}-\lambda^{-n}_{\nu\eta})|n\rangle,
\end{eqnarray}
where $\sigma=\mathrm{e}^{\mathrm{i}\frac{2\pi}{3}}$ and  
$|\chi^{\nu\eta}\rangle$ is the $\eta$-th eigenstate of the effective chain Hamiltonian $\hat{H}_{\mathrm{eff}}^\nu$ in (\ref{eqHamEff}). 
Starting from an initial wave package on the first chain with wave function $|\psi_0\rangle=\sum_n\phi^0_n|1,n\rangle=P_1|\chi^{0}\rangle$, the time evolution can be easily computed by
\begin{eqnarray}
	\vert \psi(t) \rangle  &=& \sum_{\nu\eta} \mathrm{e}^{-\mathrm{i}\epsilon_{\nu\eta} t}
	\vert \psi^{\nu\eta} \rangle \langle \psi^{\nu\eta} \vert \psi_0 \rangle 
	=\sum_lP_l\sum_{\nu}\frac{1}{3}\sigma^{\nu (l-1)}|\chi^{\nu}(t)\rangle,
	\label{EqDecompose}
\end{eqnarray} 
where $|\chi^{0}\rangle=\sum_n\phi^0_n|n\rangle$ is the corresponding initial state for effective chains.
Note that  $|\chi^{\nu}(t)\rangle= e^{-i\hat{H}_{\mathrm{eff}}^\nu t}|\chi^{0}\rangle=\sum_{\eta} \mathrm{e}^{-\mathrm{i}\epsilon_{\nu\eta} t}|\chi^{\nu\eta}\rangle\langle\chi^{\nu\eta}|\chi^{0}\rangle$ describe time evolutions governed by effective chain Hamiltonians $\hat{H}_{\mathrm{eff}}^\nu$, but starting from the same initial state $|\chi^{0}\rangle$. 
The equation (\ref{EqDecompose}) is significant, which shows that the evolving wave function on one chain of the Y-junction graph equals a superposition of evolving wave functions on three effective chains. 
The chiral quantum walk on the Y-junction graph turns into chiral quantum walks on effective chains with boundaries, or scatterings of a wave package by potential-shifted boundaries of chains.

Since our focus is on the symmetric, asymmetric, and complete transport of a wave packet on the Y-junction graph, we will analyze the wave function and total density of particles on each chain after they interact with the junction or are reflected by the edges.
To do this, we need to take into account the scattering of the initial state $\vert \chi^{0}\rangle$ by the potential-shifted boundaries of the effective chains.
As explained in detail in Section \ref{secScattering} using lattice scattering theory, the wave packet after one scattering or reflection retains the same form as the unscattered wave packet, but obtains an overall phase shift.
Specifically,
\begin{equation}
	|\chi^{\nu}(t)\rangle= \mathrm{e}^{-\mathrm{i}\hat{H}_{\mathrm{eff}}^\nu t}|\chi^{0}\rangle
	=\mathrm{e}^{\mathrm{i}\delta_\nu} \mathrm{e}^{-\mathrm{i}\hat{H}_0t}|\chi^{0}\rangle,
	\label{EqStateChian}
\end{equation}
where $\hat{H}_0=\hat{H}_\mathrm{eff}^\nu(\omega_\nu=0)$ is the ``free" Hamiltonian.  This result can be directly obtained from (\ref{eqFinal}) by noticing that $ \vert \psi_1 \rangle = \mathrm{e}^{-\mathrm{i}\hat{H}_{\mathrm{eff}}^\nu t}|\chi^{0}\rangle$ and $ \vert \psi_2 \rangle = \mathrm{e}^{-\mathrm{i}\hat{H}_0t}|\chi^{0}\rangle$ are both normalized wave functions with overlap $\vert \mathrm{e}^{\mathrm{i}\delta_\nu}\vert = 1$, then according to the condition for  holdding the equality in the Cauchy-Schwarz inequality $\langle \psi_1 \vert \psi_2 \rangle^2 \leqslant \langle \psi_1 \vert \psi_1 \rangle \langle \psi_2 \vert \psi_2 \rangle$, we arrive at (\ref{EqStateChian}).
The phase-shift in (\ref{EqStateChian}) reads
\begin{equation}
	\delta_\nu=2k_0-2\arctan\frac{\omega_\nu-\cos k_0}{\sin k_0}-\pi\quad (\mathrm{mod} \, 2\pi),
	\label{EqPhaseShift}
\end{equation}
where $k_0$ is the momentum of initial wave package. 
The result in (\ref{EqPhaseShift}) shows that the phase-shift only depends on the phase $\theta$ [since $\omega_\nu$ only depend on $\theta$, via (\ref{eqOmega})] and the initial momentum $k_0$. 
%
%
Substituting (\ref{EqStateChian}) into(\ref{EqDecompose}), we obtain the wave function
\begin{equation}
	\vert \psi(t) \rangle
	=\sum_lP_l\left[\sum_{\nu}\frac{1}{3}\sigma^{\nu (l-1)}\mathrm{e}^{\mathrm{i}\delta_\nu}\right]
	 \mathrm{e}^{-\mathrm{i}\hat{H}_0t}|\chi^{0}\rangle,
\end{equation}
for the chiral quantum walk on Y-junction graph. 
The superposition of phase-shifts or the summation in square brackets causes an enhancement or suppression of the wave function on each chain. 
Additionally, total densities of particles on different chains ($n_l$) defined above can be analytically computed by
\begin{equation}
	n_l=\left|\sum_{\nu}\frac{1}{3}\sigma^{\nu (l-1)}\mathrm{e}^{\mathrm{i}\delta_\nu}\right|^2,
	\label{eqNumberAna}
\end{equation}
when the evolving wave package is away from boundaries.

\begin{figure}[t]
	\begin{center}
		\includegraphics[width=1.00\textwidth]{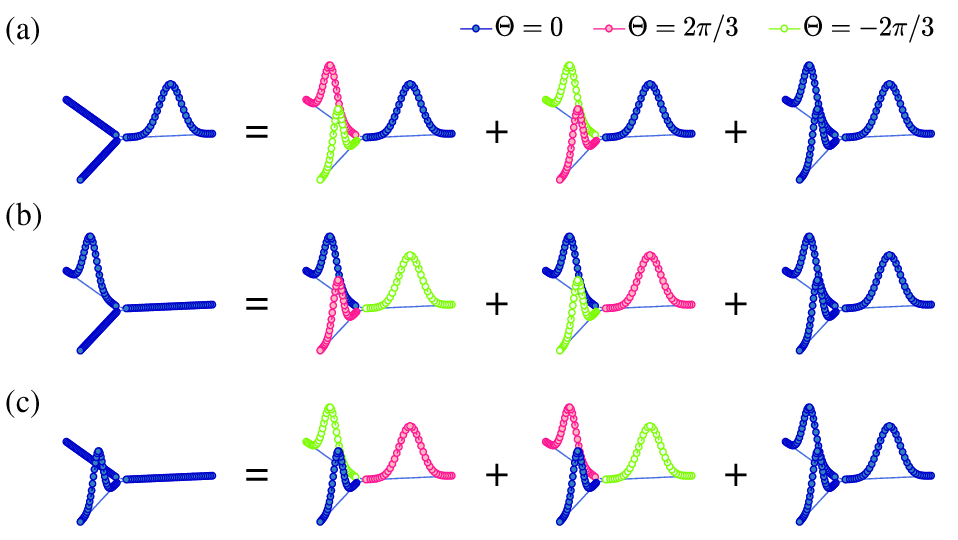}
		\caption{The schematics that explain the phyiscal mechanism behind the controlled transport phenomenon on the Y-junction showed in Figure \ref{figDynamic1}.  (a) The initial Gaussian wave function located at chain-1 can be
		decomposed into three orthogonal wave functions, which spread over the whole graph
		but with different phases on each chain, different colors refer to different phases. (b) After one scattering  the
		phases of the decomposed wave functions varies differently on each chain according to
		(\ref{eqPhaseShift}), and they compose a wave function that locates at chain-2 only. 
		(c) Similarly,
		after two scatterings, the wave package which is the superposition of the three wave
		packages, locates at chain-3 only.
		 }
		\label{figdecomposition}
	\end{center}
\end{figure}
As an example, we explain the directed complete transport phenomenon shown in Figure \ref{figDynamic1}.
When $k_0=\pi/2$ and $\theta=\pi/6$, from (\ref{eqOmega}) we obtain $\omega_\nu=-\sqrt{3} ,0, \sqrt{3}$. 
Then according to (\ref{EqPhaseShift}), the phase-shift $\delta_\nu=2\pi/3, 0, -2\pi/3$. 
Consequently, from (\ref{eqNumberAna}) we obtain 
$n_1=|(\mathrm{e}^{\mathrm{i}\cdot 2\pi/3}+\mathrm{e}^{\mathrm{i}\cdot 0}+\mathrm{e}^{-\mathrm{i}\cdot 2\pi/3})/3|^2=0$, 
which means that the superposition leads to complete destruction of the reflected wave package on the first chain. 
Similarly, 
$n_3=|(\mathrm{e}^{\mathrm{i}\cdot 4\pi/3}\mathrm{e}^{\mathrm{i}\cdot 2\pi/3}+\mathrm{e}^{\mathrm{i}\cdot 8\pi/3}\mathrm{e}^{\mathrm{i}\cdot 0}+\mathrm{e}^{\mathrm{i}\cdot 12\pi/3}\mathrm{e}^{-\mathrm{i}\cdot 2\pi/3})/3|^2=0$, 
corresponding to complete destruction of the scattered wave package on the third chain. 
However, 
$n_2=|(\mathrm{e}^{\mathrm{i}\cdot 2\pi/3}\mathrm{e}^{\mathrm{i}\cdot 2\pi/3}+\mathrm{e}^{\mathrm{i}\cdot 4\pi/3}\mathrm{e}^{\mathrm{i}\cdot 0}+\mathrm{e}^{\mathrm{i}\cdot 6\pi/3}\mathrm{e}^{-\mathrm{i}\cdot 2\pi/3})/3|^2=1$, 
meaning a complete transport of the wave package from the first to the second chain. For further clarification, we show the schematics for the calculation discussed above in Figure \ref{figdecomposition}.

For a general $\theta$, instead of the directed complete transport, (a)symmetric transport happens.
In Figure \ref{figTheta}, we present analytical results (lines) of total densities of particles on different chains $n_l$ vs. the phase $\theta$ [shown in (\ref{eqNumberAna})], for chiral quantum walks starting from the initial wave package with $k_0=\pi/2$, which agree well with the numerical simulations (dots).

\begin{figure}[t]
	\begin{center}
		\includegraphics[width=1.00\textwidth]{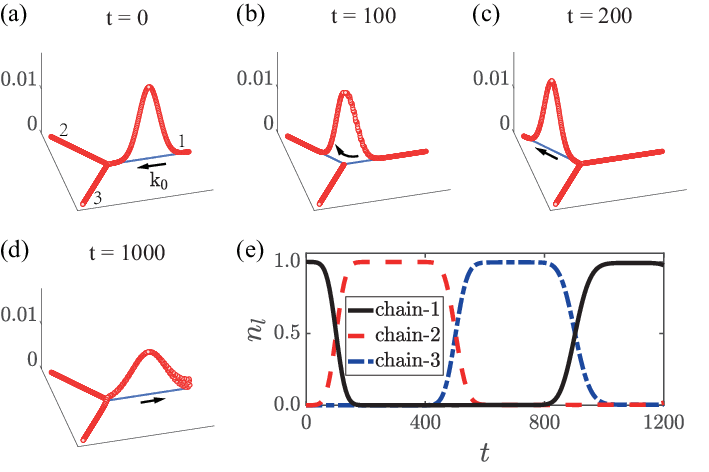}
		\caption{(a-d) Snapshots of the chiral quantum walk, from the initial wave package with momentum $k_0 = \pi/6$, and on the Y-junction graph with $N = 200$ and $\theta = 5\pi/18$. 
			The transport of wave package is directed and complete, given the condition in  (\ref{eqRelationKTheta}) is satisfied.
			The group velocity of wave package is $v_\mathrm{g} = 2\sin(\pi/6) = 1$, and it is deforming while propagating.
			(e) Corresponding time evolutions of total densities of particles on different chains.
		 }
		\label{figDynamic3}
	\end{center}
\end{figure}

The same as discussed in Section \ref{secDirected}, above we chose  $k_0 = \pi/2$ for the purpose of minimizing deformation of the evolving wave package.
However, $k_0 = \pi/2$ and $\theta = \pm\pi/6$ are not necessary conditions for the directed complete transport. 
The sufficient and necessary condition is that the relative phases among phase-shifts in (\ref{EqPhaseShift}) equal $\pm 2\pi/3$ and $0$.
In this case, the superposition on one chain is enhanced, while on the other two chains they are completely destroyed.
This gives the general condition for directed complete transport, i.e.
\begin{equation}
	3\theta + k_0 = \pi  \quad (\mathrm{mod} \,\, 2\pi).
	\label{eqRelationKTheta}
\end{equation}
The equation (\ref{eqRelationKTheta}) is the key result of this work, it relates parameter $k_0$ which is the momentum of the initial wave package and the parameter $\theta$ which decides the geometric feature of the Y-junction. Both of these parameters can impact the transport characteristics. Notice that the condition holds except for $k_0 \to 0 \,(\mathrm{mod} \, \pi)$, because when $k_0 \to 0$ expansion of the wave package is much faster than the motion of the center of mass.
According to this result, given an initial wave package with momentum $k_0$, proper $\theta$ always exist for the directed complete transport.  
We present an example in Figure \ref{figDynamic3}, corresponding to the chiral quantum walk from an initial wave package with momentum $k_0 = \pi/6$ and on the Y-junction graph with phase $\theta = 5\pi/18$. 
Main differences between chiral quantum walks shown in Figures \ref{figDynamic2} and Figure \ref{figDynamic3} are that in the latter the group velocity of wave package is $v_\mathrm{g} = 2\sin (\pi/6) = 1$, instead of $2$ in the former, and it is deforming while propagating.

\begin{figure}[t]
\begin{center}
\includegraphics[width=1.00\textwidth]{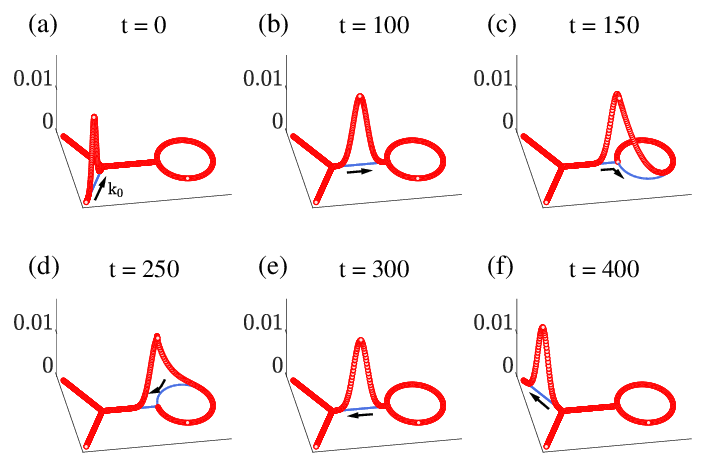}
\caption{Snapshots of the chiral quantum walk on the graph formed by combining the Y-junction graph with a circle. 
	We choose $N = 200$ for both the Y-junction graph and the circle, and the total number of sites is $4N=800$. 
	The initial Gaussian wave package is on chain-3 and with momentum $k_0 = \pi/2$. 
	We choose the phase $\theta = \pi/6$ to ensure the condition in (\ref{eqRelationKTheta}) for achieving directed complete transport.
}
\label{figDynamic4}
\end{center}
\end{figure}

The above discussion about the directed complete transport can be more general. 
%
%
The simple relation in  (\ref{eqRelationKTheta}) enables us to construct more general and complicated graphs, which support the directed complete transport. 
For example, in Figure \ref{figDynamic4} we show the directed complete transport on a graph, formed by combining a Y-junction graph with a circle.
The end site of chain-1 and two neighboring sites of the circle form a triangle with edge-weights $\mathrm{e}^{\pm \mathrm{i}\theta}$, same as in Figure \ref{figSchematic1}. 
An even more profound example is the binary tree graph, formed by connecting ends of chains for multiple copies of the Y-junction graph. 
One can implement quantum versions of binary tree search algorithms by controlling the phase $\theta$. 
For instance, evolving from an initial wave package with momentum $k_0=\pi/2$, the depth-first binary tree search can be achieved in chiral quantum walks, by setting the phase $\theta=\pm\pi/6$. 
And the left side child will be visited first when $\theta=\pi/6$, while when $\theta=-\pi/6$ the right side child will be the first. 
On the other hand, the breadth-first binary tree search can be achieved by setting the phase $\theta=0$. 
These results are in support the conclusion that the Y-junction structure under proper parameters can server as diverting valves, and the wave package can be manipulated to completely transport onto one particular chain of the Y-junction graph.

\section{Scattering theory on lattice} \label{secScattering}

Scattering theory is one of the most important foundations in modern physics \cite{newton2002,faddeev2016,weinberg1964}.  
While the relativistic one was well developed and broadly applied in high energy physics \cite{iliopoulos2021}, the non-relativistic quantum scattering theory for continuous systems has been a core tool in studying ultra-cold atomic physics \cite{taylor1972, chin2010}. 
However, the quantum scattering theory for lattice models was rarely discussed \cite{kroger1990, feldman2004, rubtsova2010}, for two possible reasons: 1) in the presence of lattice, energy band, Brillouin zone, and discrete nature of lattice prevent us from using the analytical tools developed in the traditional scattering theory; 2) the $t\rightarrow\pm\infty$ limits of the Green's function or M\o ller operator $\Omega(t)= \hat{U}(-t)\hat{U}_0(t)$, which are vital in the traditional scattering theory, are not well defined, due to the presence of boundaries. 
Traditionally, the limit of Green's function is computed perturbatively by summing series of the free Green's function. 
However, here we compute finite time Green's function directly, based on the exact solvability of the effective chain Hamiltonians in  (\ref{eqHamEff}).

\begin{figure}[b]
	\begin{center}
	\includegraphics[width=0.8\textwidth]{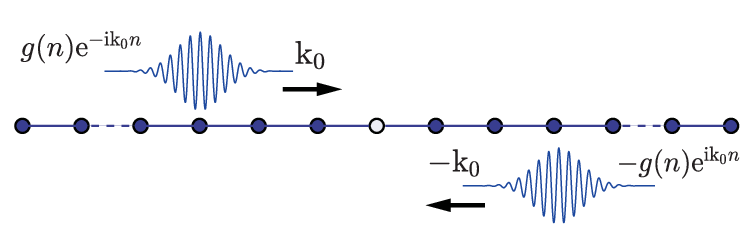}
	\caption{Schematics, which presents doubling of the $N$-site chain 
	to the $(2N+1)$-site chain. 
	The chiral quantum walk on the $N$-site chain with open boundaries is equivalent to the time evolution on $(2N+1)$-site chain, with odd parity wave functions to ensure the zero amplitude on central site. }
	\label{figSchematic2}
	\end{center}
\end{figure}

\begin{figure}[tb]
	\begin{center}
		\includegraphics[width=1.00\textwidth]{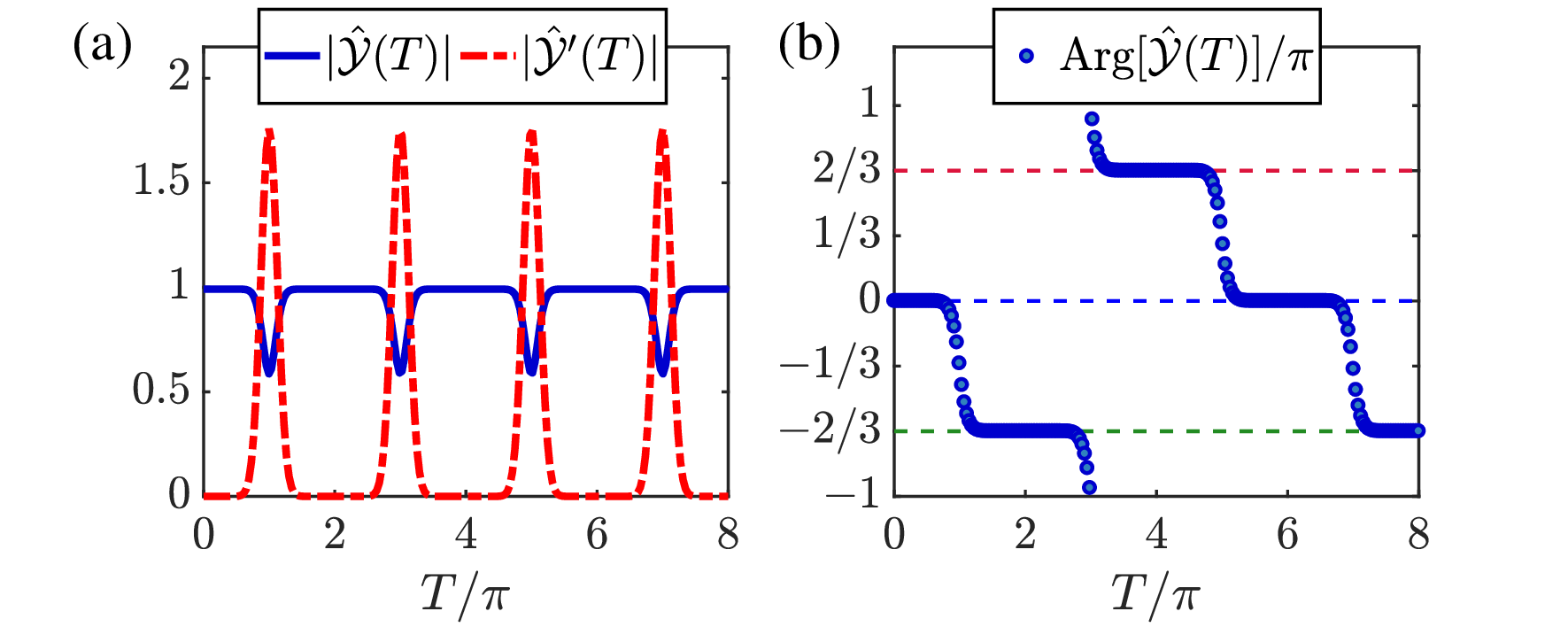}
		\caption{Typical behaviors of the functions $\hat{\mathcal{Y}}(T)$
			and $\hat{\mathcal{Y}}^\prime(T)$. 
			We choose an initial Gaussian wave package with momentum $k_0 = \pi/2$, a chain with size $N=200$, and the boundary potential $\omega = \sqrt{3}$ (corresponding to phase $\theta = \pi/6$). 
			$|\cdot|$ and $\mathrm{Arg[\cdot]}$ denote the magnitude and amplitude of a complex number, respectively. 
			At the time (horizontal axis), where the magnitude and amplitude dramatically change, the wave package is colliding with potential-shifted boundaries.  }
		\label{figYt}
	\end{center}
\end{figure}

Before computing Green's function, we first study the time evolution of a wave package on a uniform lattice with open boundaries. 
Let $\hat{H}_0=\hat{H}_\mathrm{eff}^\nu(\omega_\nu=0)$ be the Hamiltonian of a potential-free chain with $N$ sites, and we denote eigenvalues and corresponding eigenstates of it by $E_p$ and $|p\rangle$ respectively, i.e. $\hat{H}_0|p\rangle=E_p|p\rangle$. 
We obtain that
\begin{equation}
	\left(\begin{array}{c}
		 \\
		\hat{H}^{2N+1}_0\\
		 \\
	\end{array}
		 \right)\left(\begin{array}{c}|p\rangle\\
	0\\
	-|p\rangle
\end{array}\right)=E_p\left(\begin{array}{c}|p\rangle\\
0\\
-|p\rangle\end{array}\right),
\label{EqParity}
\end{equation}
where $\hat{H}^{2N+1}_0$ is the same as $\hat{H}_0$, but for a $(2N+1)$-site chain. 
This means that the propagation of a wave package and the reflection of it by a boundary are mapped to the propagation on a chain with $(2N+1)$ sites, and  the wave function on the new chain has odd parity [see Figure \ref{figSchematic2}]. 
To include boundary reflection, we first double the $N$-site chain, and then between them we insert a site, whose amplitude of wave function is always demanded zero to represent the original open boundary.    
Supposing that the wave function of initial wave package on the original $N$-site chain is $\phi^0_n = g(n)\mathrm{e}^{-\mathrm{i} k_0n}$ with $g(n)$ the envelope and $k_0$ the momentum, when mapped onto the $(2N+1)$-site chain, a wave package with momentum $-k_0$ and an additional minus sign are added to the right half of chain, to ensure the odd parity or the zero amplitude on central site  [see also Figure \ref{figSchematic2}].   
Thus, after colliding with the open boundary of $N$-site chain, a wave package turns into its parity-symmetric one with an inverse momentum $-k_0$ and an additional minus sign (or phase-shift $\pi$). 
Furthermore, when a wave package moves back and forth on the $N$-site chain, correspondingly it evolves on an enlarged chain formed by unfolding enough times, accompanying an inversion of momentum and phase-$\pi$ shift after every time it collides with a boundary. 
Particularly, known from Section \ref{secDirected}, when $k_0=\pi/2$, the wave package keeps its envelope invariant on the unfolded chain, and only the inversion of momentum and the phase-$\pi$ shift left.
Now, we study the time evolution governed by $\hat{H}^\nu_\mathrm{eff}$, with reference to the one governed by $\hat{H}_0$ (presented above),
we compute the Green's function
\begin{equation}
\mathcal{Y}(t) = \langle \psi_0 \vert  
\mathrm{e}^{\mathrm{i}\hat{H}_0 t} 
\mathrm{e}^{-\mathrm{i}\hat{H}_\mathrm{eff}^\nu t} 
\vert \psi_0 \rangle,
\label{eqDefYt}
\end{equation}
where $\vert \psi_0 \rangle$ denotes the initial state. 
In the following we take the initial Gaussian state $|\psi_0\rangle=(2\pi\sigma^2)^{-1/2}\sum_n\mathrm{e}^{-(n-n_0)^2/2\sigma^2-\mathrm{i}k_0n}|n\rangle$ as an example, and drop scripts `$\nu$' and `$\mathrm{eff}$' without ambiguity. 
We denote $\vert q \rangle$ as the $q$-th eigenstate of $\hat{H}$ with eigenvalue $\epsilon_q$, and $\vert p \rangle$ as the $p$-th eigenstate of $\hat{H}_0$ with eigenvalue $E_p$.
Note that for simplicity we use letters $p$ and $q$ to denote vectors of different Hamiltonians, and in general $\vert p = z \rangle \neq \vert q = z \rangle$ for $ z \in \mathbb{N}$. 
The letter $n$ is used to denote the Wannier base $\vert n \rangle$.
Based on eigen-decomposition, $\mathcal{Y}(t)$ can be rewritten as
\begin{equation}
	\mathcal{Y}(t) = \langle \psi_0 \vert  
	\mathrm{e}^{\mathrm{i}\hat{H}_0 t} 
	\mathrm{e}^{-\mathrm{i}\hat{H} t} 
	\vert \psi_0 \rangle, = \sum_{p,q} \langle \psi_0 \vert p \rangle
	\langle p \vert q \rangle
	\langle q \vert \psi_0 \rangle
	\mathrm{e}^{\mathrm{i} t (E_p - \epsilon_q)}.
	\label{EqYtExpansion} 
\end{equation}
Following the discussion in Section \ref{secEigen}, eigenstates are written as
\begin{equation}
	\vert p \rangle = \frac{1}{\Xi(\theta_p)}\sin (\theta_p n)|n\rangle ,  \quad 
	\vert q \rangle = \frac{1}{\Xi(\theta_q)}\sin (\theta_q n)|n \rangle,
	\label{EqEigenstate}
\end{equation}
where $\theta_p$ ($\theta_q $) is the solution of equation $\sin [(N+1)\theta ]/ \sin [N \theta ]= 0 (\omega)$ in the interval $(\frac{\pi}{N}(p(q)-1), \frac{\pi}{N}p(q))$, and $\Xi(\theta)=\frac{1}{2}\sqrt{2N+1-\sin[(2N+1)\theta]/\sin \theta}$ is a normalization. 
Then, for the initial Gaussian wave package, we have
\begin{eqnarray}
	\langle q \vert \psi_0 \rangle	= \frac{		\mathrm{e}^{-\mathrm{i}(k_0-\theta_q)n_0}\mathcal{\theta}_3(\mathrm{e}^{-\frac{1}{2\sigma^2}}, \frac{k_0-\theta_q}{2})-\mathrm{e}^{-\mathrm{i}(k_0+\theta_q)n_0}\mathcal{\theta}_3(\mathrm{e}^{-\frac{1}{2\sigma^2}}, \frac{k_0+\theta_q}{2})}{2 \mathrm{i} \mathcal{N} \Xi(\theta_q)},
	\label{EqOverlap}
\end{eqnarray}
where $\theta_3$ is the Jacobi-theta function and $\mathcal{N} = \sqrt{\theta_3(0, \mathrm{e}^{-1/\sigma^2})}$ is the normalization of Gaussian package. 
The overlap $\langle p \vert \psi_0 \rangle$ is the same as the equation (\ref{EqOverlap}), but with a replacement of $q$ by $p$. 
Since $\sigma \sim O(N) \gg 1$, $\theta_3$ in (\ref{EqOverlap}) is an asymptotically narrow Gaussian function of variables $k_0-\theta_q$ and $k_0+\theta_q$, with a period $2\pi$. 
Moreover, the second term in numerator can be ignored, since $k_0 + \theta_q$ never closes to $2\pi$. 
According to (\ref{eqPolyEq}), $\theta_p$ and $\theta_q $ can be estimated by
\begin{eqnarray}
	\theta_p &&= \frac{\pi}{N}(p-1) + \frac{1}{N}f( \frac{\pi}{N}p,0) + O(\frac{1}{N^2}),
	\nonumber \\
	\theta_q &&= \frac{\pi}{N}(q-1) + \frac{1}{N}f( \frac{\pi}{N}q,\omega) + O(\frac{1}{N^2}),
	\label{eqThetaResults}
\end{eqnarray}
with the analytical function
\begin{equation}
	f(x,\omega) = \frac{\pi}{2} - \arctan \frac{\omega - \cos x}{\sin x}, \quad \omega \in \mathbb{R} \quad \mathrm{and} \quad x \in (0, \pi).
	\label{eqf}
\end{equation} 
Now we obtain the major ingredient in (\ref{EqYtExpansion})
\begin{equation}
	\langle q \vert \psi_0 \rangle \langle \psi_0 \vert p \rangle	= \frac{\pi  \sigma^2}{\mathcal{N}^2N}\mathrm{e}^{\mathrm{i}\frac{n_0 \pi}{N}(q-p-\beta/\pi)}	\mathrm{e}^{-\frac{\pi^2\sigma^2}{2N^2}[(q-N\frac{k_0}{\pi})^2 + (p-N\frac{k_0}{\pi})^2]},
	\label{eqCalculationOverlap3}
\end{equation}
the fact $\Xi(\theta) = \sqrt{\frac{N}{2}} + O(1)$ has been used above and phase $\beta$ is defined by $\beta\equiv f(k_0, 0) - f(k_0, \omega)$. 
The above equation indicates that main contributions of the summations in (\ref{EqYtExpansion}) come from terms around $q, p = N\frac{k_0}{\pi}$, since they decay to zero very fast when the integer $q$ or $p$ moves away. 
In the same spirit, we learn that around $q, p = N\frac{k_0}{\pi}$, one has
\begin{equation}
	\langle p \vert q \rangle 	\approx \frac{1}{2N}\frac{\sin \frac{2N+1}{2}[\frac{\pi}{N}(p-q)+\frac{\beta}{N}]}	{\sin \frac{1}{2}[\frac{\pi}{N}(p-q) + \frac{\beta}{N}]} 	\approx -\frac{(-1)^{q-p}}{\pi} \cdot \frac{\sin \beta}{q-p-\frac{\beta}{\pi}}.
	\label{eqCalculationOverlap4}
\end{equation}
Moreover, by Taylor expansion of the spectra around $q, p \to N\frac{k_0}{\pi}$, we obtain
\begin{equation}
	\mathrm{e}^{\mathrm{i}t(E_p - \epsilon_q)}
	\approx \mathrm{e}^{\mathrm{i}t\cdot 2 \sin k_0 \cdot \frac{\pi}{N}(q-p-\frac{\beta}{\pi})}.
	\label{eqCalculationOverlap5}
\end{equation}
Substituting above three equations into (\ref{EqYtExpansion}), we finally obtain
\begin{equation}
	\hat{\mathcal{Y}}(T(t))  \triangleq \mathcal{Y}(t) 
	= \sum_{q,p} \frac{\sigma^2}{2\mathrm{i}\mathcal{N}^2N}\cdot
	\frac{1 - \mathrm{e}^{2\mathrm{i}\beta}}{q-p-\frac{\beta}{\pi}}\cdot
	\mathrm{e}^{\mathrm{i}(q-p-\frac{\beta}{\pi})T(t)-\frac{\pi^2\sigma^2}{2N^2}(q^2 + p^2)},
	\label{eqYtFinal}
\end{equation} 
where
\begin{equation}
T(t) = \pi + \frac{n_0 \pi}{N} + \frac{2\pi t \sin k_0}{N}.
\end{equation}
The linearity between $T$ and $t$ implies that the group velocity of wave package is simply $v_\mathrm{g}=2\sin k_0$. 
Note that, $p$ and $q$ take integers in (\ref{eqYtFinal}), thus $\hat{\mathcal{Y}}(T)$ possesses quasi periodicity described by
\begin{equation}
	\hat{\mathcal{Y}}(T + 2\pi) = \mathrm{e}^{\mathrm{i}\delta}
	\hat{\mathcal{Y}}(T),
	\label{eqPeriodicYt}
\end{equation}
where $\delta$ is the most important phase-shift
\begin{equation}
	\delta = 2k_0 - 2 \arctan \frac{\omega - \cos k_0}{\sin k_0} - \pi.
	\label{eqPhaseShift}
\end{equation}
The summation in (\ref{eqYtFinal}) is hard to analytically compute, however, the derivative of $\hat{\mathcal{Y}}(T)$ can be computed analytically by
\begin{equation}
	\hat{\mathcal{Y}}^\prime(T) = \frac{\sigma^2(1-\mathrm{e}^{2\mathrm{i}\beta})}
	{2\mathcal{N}^2N}\mathrm{e}^{-\mathrm{i}T\frac{\beta}{N}}
	[\theta_3 (T/2, \mathrm{e}^{-\frac{\pi^2\sigma^2}{2N^2}})]^2.
	\label{eqYtdt}
\end{equation}
Typical behaviors of the functions $\hat{\mathcal{Y}}(T)$ and $\hat{\mathcal{Y}}^\prime(T)$ are presented in Figure \ref{figYt}.

Now we restrict on the first collision with the potential-shifted boundary. 
Before hitting the boundary, we have $\hat{\mathcal{Y}}(T) =1$, since $\hat{H}_0$ and $\hat{H}_\mathrm{eff}^\nu$ are the same when away from boundaries. 
After reflection we have $\hat{\mathcal{Y}}(T) =\mathrm{e}^{\mathrm{i}\delta}$ from the quasi periodicity. Thus
\begin{equation}
\langle \psi_0 \vert  
	\mathrm{e}^{\mathrm{i}\hat{H}_0 t} 
	\mathrm{e}^{-\mathrm{i}\hat{H}_\mathrm{eff}^\nu t} 
	\vert \psi_0 \rangle=\mathrm{e}^{\mathrm{i}\delta_\nu}, 
	\label{eqFinal}
\end{equation}
when the wave package is away from boundaries.
This equation directly leads to (\ref{EqStateChian}).

\section{Conclusion and discussion}

In this paper, we examine the asymmetric and directed transport of wave packets on chiral chains and Y-junction graphs. We prove that quantum walks on chiral chains are equivalent to those on non-chiral chains, but with additional momenta in the initial wave packets. This explains the novel transport phenomenon observed in previous studies. We then analyze the chiral quantum walks of wave packets with finite initial momenta on the Y-junction graph, where phases in the hopping on the Y-junction triangle break time-reversal symmetry. The asymmetric and directed transport is fully controlled by the phase.
We provide an analytical explanation for these novel phenomena by showing that the chiral quantum walk on the Y-junction graph can be viewed as a superposition of three wave packets that evolve on three effective open chains with potential-shifted boundaries. We then use scattering theory on the lattice to determine the phase shift of a wave packet after it collides with the boundary and explain how the phase shifts result in the asymmetric and directed transport. The conditions for directed transport are also derived.
Our findings have potential applications in the development of quantum binary tree search algorithms. Future studies of transport in chiral DTQWs on graphs \cite{zhou2021,burgarth2013,cameron2014,mulken2011,	manouchehri2014,izaac2017,chen2017} will also be of interest, and our scattering theory on the lattice will be valuable.
%
%
%

%
The physics discussed in the paper can be experimentally tested in several setups. In ultracold atoms, programmable effective hoppings and geometries have been reported using Raman-addressing techniques \cite{periwal2021, hung2016, jaksch2003, bermudez2011, kolovsky2011}. For example, the two-dimensional Hofstadter Hamiltonian was realized in this manner \cite{aidelsburger2013a,miyake2013}, whose one-dimensional projection is precisely the chain Hamiltonian in (\ref{eqHamiltonian}). Furthermore, a few works have attempted to experimentally realize chiral quantum walks on directed graphs via complicated quantum gate networks \cite{wu2020, wang2020}, but they have limited system sizes. Maybe, the most promising platform for simulating this physics is the photonic waveguides \cite{tang2018,papadopoulos2000},  in which each site of the chains or graphs corresponds to a waveguide and the graph geometry can be easily adjusted. In the tight-binding limit, classical light propagation among the sites is governed by an equation similar to the Schr\"{o}dinger equation. The hopping amplitude between different waveguides is represented by $t_{j,l}$, and the light intensity at site $j$ is represented by $|\phi_j|^2$.
%

%

\ack{This work is supported by the National Natural Science Foundation of China under grant No.12274419, No.12134015 and No.12175290, and the CAS Project for Young Scientists in Basic Research under grant No.YSBR-055. }

\vspace{10pt}




\begin{figure}[t]
	\begin{center}
		\includegraphics[width=1.0\textwidth]{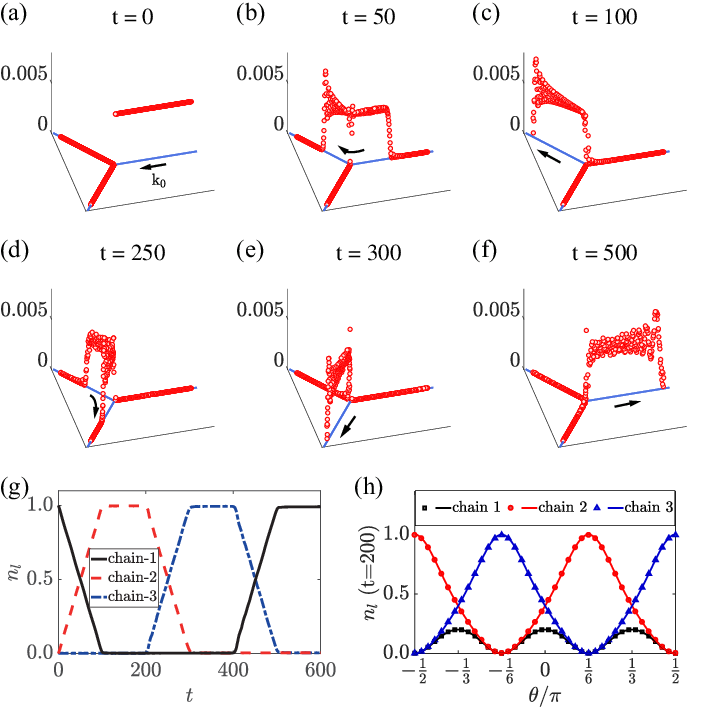}
		\caption{(a-f) Snapshots of the chiral quantum walk, from the initial square wave package with momentum $k_0 = \pi/2$, and on the Y-junction graph with $N = 200$ and phase $\theta = \pi/6$. 
			Same as in Figure \ref{figDynamic1}, scattered by the Y-junction the wave package is completely transported onto the right hand side chain, and the group velocity of wave package is also $v_\mathrm{g} = 2\sin (\pi/2) = 2$. 
			(g) Corresponding time evolutions of total densities of particles on different chains. 
			(h) Total densities of particles on different chains $n_l$ vs. phase $\theta$, of the wave package after the first collision with the Y-junction. 
		 }
		\label{figAppendix}
	\end{center}
\end{figure}

\section*{Appendix: Chiral quantum walks from the initial square wave package} \label{secApp}

Notice that in derivation of key results in (\ref{EqPhaseShift}) and (\ref{eqRelationKTheta}), there is no special prerequisite on the initial wave package, except that in momentum space the wave package is a narrow function centering on a momentum $k_0$ which satisfies $k_0 \neq n\pi $, $\forall n \in \mathbb{Z} $. 
Therefore, the directed complete transport similarly occurs in the chiral quantum walk from an initial square wave package, which is of great interest in the study of quantum search algorithms. 
In Figure \ref{figAppendix}, we show the directed complete transport in the chiral quantum walk on the Y-junction graph and from the initial square wave package $\vert \psi_0 \rangle \sim \mathrm{e}^{\mathrm{i}k_0n}|1,n\rangle$. 
Despite suffering from notable finite-size effects, the propagation of a square wave package obeys the same rules as the Gaussian one.

\vspace{10pt}

\bibliographystyle{unsrt}

\end{document}